# Social Bayesian Learning in the Wisdom of the Crowd

Dhaval Adjodah, Yan Leng, Shi Kai Chong, Peter Krafft, Alex Pentland

February 28, 2017

## 1 Introduction

Being able to correctly aggregate the beliefs of many people into a single belief is a problem fundamental to many important social, economic and political processes such as policy making, market pricing and voting. Although there exist many models and mechanisms for aggregation, there is a lack of methods and literature regarding the aggregation of opinions when influence and learning between individuals exist. This is in part because there are not many models of how people update their belief when exposed to the beliefs of others, and so it is hard to quantify the dependencies between people's mental models which is essential to minimizing redundancies in the aggregation. In this paper, we explore many models of how users influence and learn from each other, and we benchmark our models against the well-known DeGroot model [1]. Our main contributions are: 1) we collect a new dataset of unprecedented size and detail to be posted online; 2) we develop a new Social Bayesian model of how people update their mental models, 3) we compare of our model to other well-known social learning models. Specifically, we show that our new Social Bayesian model is superior to the other models tested.

## 2 Literature Review

There is evidence showing that individuals do not always give their best guess when answering a question, but instead sample from an internal distribution [8, 5], which explains why asking the same question multiple times improves guesses. The problem of aggregation then becomes an exercise of combining people's individual distributions into a group distribution. Although there is an extensive literature on how to combine distributions [6, 3, 7], there is limited work on how to combine such distributions when there is influence and learning between individuals. This is an important distinction because if people have correlated beliefs, the aggregation over distributions should incorporate the decrease in entropy that occurs from having shared beliefs compared to the ideal (and often assumed) case when people have independent beliefs. We build on the work of [2, 4] to model the individual cognitive process of individuals as they update their beliefs after being exposed to social information. These individual distributions could then be used to calculate inter-individual correlations in beliefs, which can then be used to aggregate with correlation. In this paper, we focus on how to estimate people's individual belief distribution.

## 3 Data

Over the span of 6 months, we ran 7 sequential and independent Wisdom of the Crowd (WoC) rounds of 2037 students (from several online classes) making 17420 price predictions of real financial assets (e.g. the S&P 500) over a period for 3 weeks each round. We collaborated with the crowdsourced stock rating website Vetr.com and designed the prediction process as follows: students go to our page and make a (pre-social) prediction of an instrument's price, then they are showed a histogram of their peer's predictions after which they make another (post-social) prediction. They have to provide confidence ratings and answers to surveys relating to their past experience in finance. Because the students are from a class, we also have their demographics, grades, and discussion forum data. Interestingly, we also have two rounds of prediction that happened during extraordinary market environments: the Brexit vote and Trump's election.

## 4 Analysis & Results

Our new Social Bayesian model is based on the fact that we know each student's pre-social and post-social predictions, and the exact histogram (of their peers' predictions) they are exposed to. This model updates individual belief about the price of an asset using social information as follows:

$$P(post|SI, prior) = \frac{P(SI, prior|post)P(post)}{P(SI, prior)}$$
$$= \frac{P(SI|post)P(prior|post)P(post)}{P(SI, prior)}$$
$$= \frac{\frac{P(post|SI)P(SI)}{P(post)} \frac{P(post|prior)P(prior)}{P(post)} P(post)}{P(SI, prior)}$$
$$= \frac{P(post|SI)P(SI)P(post|prior)P(prior)}{P(post)P(SI, prior)}$$
$$= \kappa \frac{P(post|SI)P(post|prior)}{P(post)}$$

where $\kappa$ is a constant, $\kappa = \frac{P(SI)P(prior)}{P(prior,SI)}$, because all terms in $\kappa$ are constant and known given *prior* and *SI*. *post* is the post-social prediction, *prior* is the pre-social prediction and *SI* is the mean of the social information. We only use Bayes theorem in this derivation. The only assumption we make (between the first and second step) is that *SI* and *prior* are conditionally independent on *post*. Because there are potential exogenous and indirect dependencies, we test the fit of our model under this assumption. We do so by numerically calculating the joint distributions: we first construct distributions from point estimates (each person's predictions) and we then discretize the distributions into several bins. We then calculate the Mean Absolute Error of how our model's post-social prediction compares to the individual's actual post-social prediction. As we can see in Table 1, our Social Bayesian model outperforms the DeGroot model baseline. Additionally, we test our models against another class of models we built which are loosely Naive Bayesian [9] (because the likelihood is assumed to be the social information seen by a user) using 1) the mean or mode of the posterior distribution for the post-social prediction, 2) either an assumed normal or an empirical distribution for the likelihood (social histogram), and 3) either a normal or uniform distribution for the prior. Finally, we also try a simple probabilistic learning model where $P(post) = P(SI \cap prior) = P(SI) * P(prior)$.

Table 1: Model Comparison (all values are percentages)

|  | Mean Absolute Error over Rounds | | | | | | |
|---|---|---|---|---|---|---|---|
|  | 1 | 2 | 3 | 4 | 5 | 6 | 7 |
| Normal Approx. | 2.79 | 6.03 | 2.21 | 2.10 | 1.42 | 2.73 | 2.75 |
| Em_Mean_Norm | 3.37 | 7.31 | 2.63 | 2.61 | 1.79 | 3.48 | 3.51 |
| Em_Mean_Uni | 3.94 | 7.39 | 2.87 | 2.41 | 1.79 | 3.04 | 3.49 |
| Em_Mode_Norm | 3.38 | 7.40 | 2.61 | 2.62 | 1.79 | 3.48 | 3.48 |
| Em_Mode_Uni | 3.30 | 6.62 | 2.49 | 2.51 | 1.60 | 2.89 | 3.37 |
| DeGroot | 2.51 | 5.27 | 1.94 | 1.86 | 1.24 | 2.62 | 2.29 |
| Prob. Learning | 2.05 | 5.23 | 1.97 | 1.69 | 1.21 | 2.47 | 2.32 |
| Social Bayesian | 1.52 | 5.13 | 1.92 | 0.82 | 0.63 | 1.28 | 0.86 |
| **Improvement** | **54.2** | **10.5** | **2.0** | **87.7** | **58.9** | **122.3** | **147.1** |

As we can see in Figure 1, our newly developed Social Bayesian learning model clearly outperforms all the empirical models used in the literature in all rounds. We improve 54.2%, 10.5%, 2.0%, 87.7%, 58.9%, 122.3% and 147.1% over the best base models. The improvement is calculated using $improv = \frac{error_{baseline} - error_{new}}{1 - error_{baseline}}$. This indicates that our Social Bayesian model is a better approximation to how people update their belief when exposed to the beliefs of others.

Because we now also have an accurate representation of people's belief (as opposed to the usual single point-estimates in WoC contexts), we could also estimate the correlations between each user's belief distribution using, for example, the Kullback–Leibler divergence. Using such measures of dependence, one could improve the WoC through a variety of means because one could now find users with beliefs that are too similar to each other (e.g. to reduce redudancy), or to find outliers who have very different beliefs (e.g. to remove inacurrate predictors).

Thus, our Social Bayesian formuation not only outperforms other models from literature, but it also allows us to construct a representation that can be used to calculate useful inter-individual correlations for improved aggregation.

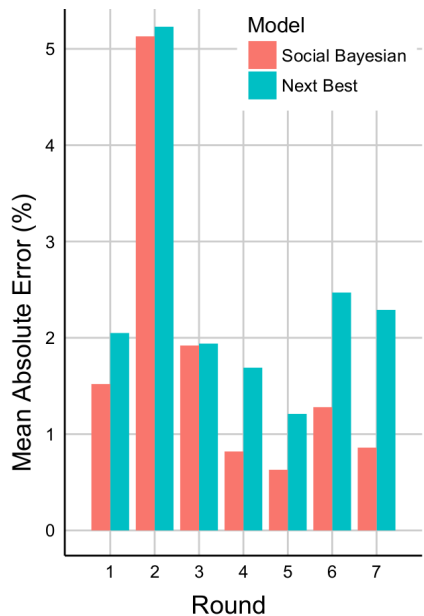

Figure 1: Mean Absolute Error for Social Bayesian model vs. next best model